\documentclass[12pt]{iopart}
\usepackage{iopams}  
\usepackage{harvard}  

\def\d {\mbox{d}}
\newcommand{\ul}[1]{\overline{#1}}
\newcommand{\ut}[1]{\tilde{#1}}
\begin{document}

\def\be{\begin{equation}}
\def\ee{\end{equation}}
\def\bea{\begin{eqnarray}}
\def\eea{\end{eqnarray}}

\title[$\gamma$ in light-Scalar-Tensor theory with a universal scalar/matter coupling]{The $\gamma$ parameter  in Brans-Dicke-like (light-)Scalar-Tensor theory with a universal scalar/matter coupling and a new decoupling scenario}

\author{Olivier Minazzoli$^{1,2}$}

\address{$^1$Jet Propulsion Laboratory, California Institute of Technology,
4800 Oak Grove Drive, Pasadena, CA 91109-0899, USA}
\address{$^2$UMR ARTEMIS, CNRS, University of Nice Sophia-Antipolis,
Observatoire de la C\^ote d’Azur, BP4229, 06304, Nice Cedex 4, France}
\ead{ominazzoli@gmail.com}
\begin{abstract}
The post-Newtonian parameter $\gamma$ resulting from a universal scalar/matter coupling is investigated in Brans-Dicke-like Scalar-Tensor theories where the scalar potential is assumed to be negligible. Conversely to previous studies, we use a perfect fluid formalism in order to get the explicit scalar-field equation. It is shown that the metric can be put in its standard post-Newtonian form. However, it is pointed out that $1-\gamma$ could be either positive, null or negative for finite value of $\omega_0$, depending on the coupling function; while Scalar-Tensor theories without coupling always predict $\gamma<1$ for finite value of $\omega_0$. Moreover, regardless the value of $\omega_0$, the subclass of theories satisfying $\gamma=1$ and $\beta=1$ surprisingly leads to almost the same phenomenology as general relativity in the solar system.
\end{abstract}

\maketitle
\section{introduction}

Brans-Dicke-like scalar-tensor theories are known to be good alternative candidates to General Relativity (GR) \cite{GEF2004,Will_book93,Will-lrr-2006-3,damourCQG92}. Similar theories with both scalar/curvature and scalar/matter couplings generically appear in (gravitational) Kaluza-Klein theories with compactified dimensions \cite{Sstring88,fujiiBOOKst}, or in string theories at the low energy limit \cite{DamPolyNPB94,damourPRD96,damourPRL02,gasperiniPRD02}. From a more phenomenological point of view, it seems that some restrictions, such as gauge and diffeomorphism invariances, single out such type of theories as well \cite{armendarizPRD02}. Recently, scalar/matter couplings have been introduced in several different type of theories:  in $f(R)$ gravity \cite{bertolamiPRD07,bertolamiPRD08,bertolamiCQG08,sotiriouCQG08,defeliceLLR13,sotiriouRMP10,harkoPRD13}, in Brans-Dicke theories \cite{dasPRD08,bisabrPRD12,moffatIJMPD12}, or in the so-called MOG (MOdified Gravity) \cite{moffatJCAP06,moffatCQG09}.
Such theories are often invoked as a possible explanation for dark Energy -- which is generically attributed to a scalar field \cite{peebleRMP03} --, for the possible observed variation of the fine structure constant in both time \cite{webbPRL2001} and space \cite{webbPRL11,murphyMNRAS03}, for (at least) some phenomena usually attributed to dark matter; or to generically predict violations of the equivalence principle \cite{DamPolyNPB94,dvaliPRL02,olivePRD08,damourPRD10,damourCQG12}.

The post-Newtonian phenomenology of theories with scalar/matter coupling has been thoroughly studied, notably in \cite{DamPolyGRG94,DamPolyNPB94,damourPRD10} with an emphasize on the composition-dependent phenomena. In this paper, we want to focus our attention on the composition-independent part of the phenomenology -- namely, on the post-Newtonian parameter $\gamma$ (see (77) in \cite{damourPRD10} for instance). To this end, we restrict ourselves to an action with a universal scalar/matter coupling, where the coupling is given by a field that couples universally to all the material fields through a function in factor of the material part of the Lagrangian. It has to be noticed that such an action satisfies the condition for the dynamical decoupling studied in \cite{DamPolyGRG94,DamPolyNPB94}. However, instead of modeling our fluids by point particles as in \cite{DamPolyGRG94,DamPolyNPB94}; we chose to use a perfect fluid formalism based on a recent result given in \cite{moi_hPRD12}, that tells that with a non-minimal coupling, basic assumptions such as the conservation of the matter fluid current imply that the Lagrangian of a perfect fluid is minus the total energy density and nothing else. 

Our result is that, depending on the scalar/matter coupling function, the post-Newtonian parameter can be less, equal or more than one  . Since this important fact has been missed in previous studies, we explain the reason why it has not been derived previously. We emphasize that the discrepancy on $
\gamma$ does not lie on the discrepancy of formalisms used (non-interactive point particles versus perfect fluid); but rather lies on a wrong property assumed in previous studies. Moreover, we show that the specific cases that lead to $\gamma=1$, lead to a new decoupling mechanism such that the scalar-field acts as if it was weakly coupled to matter fields in regions where the pressure of the gravitational bodies is significantly lower than their energy density -- such as in the solar system.

In section \ref{sec:eqm}, we derive the equations of motion coming from the considered action. Then, in section \ref{sec:magnitude} we concentrate on the post-Newtonian parameter $\gamma$ resulting from such theories. In section \ref{sec:magnitude2} we focus our attention on a subclass of theories where the scalar-field is naturally almost decoupled to the gravitational sources in the post-Newtonian regime. Finally, we give our conclusions in \ref{sec:concl}. 
\section{Equations of motion}
\label{sec:eqm}

The action describing Brans-Dicke-like theories with a universal scalar/matter coupling can be written as follows:
\begin{equation}\label{eq:action}
S=\int d^4x\sqrt{-g} \left( \Phi R -
\frac{\omega(\Phi)}{\Phi} \left(\partial_{\sigma}\Phi \right)^2+ 2f(\Phi)\mathcal{L}_m(g_{\mu \nu},\Psi) \right),
\end{equation}
where $g$ is the metric determinant, $R$ is the Ricci scalar
constructed from the metric $g_{\mu \nu}$ , $\mathcal{L}_m$ is the material Lagrangian and $\Psi$ represents the non-gravitational fields. From this action, and defining
\begin{equation}
T_{\mu \nu}=-\frac{2}{\sqrt{-g}} \frac{\delta(\sqrt{-g}\mathcal{L}_m)}{\delta g^{\mu \nu}},
\end{equation}
 one gets the following equations of motion:
\begin{eqnarray}
R_{\mu \nu}-\frac{1}{2}g_{\mu \nu}R&=& \frac{f(\Phi)}{\Phi}T_{\mu
\nu}+\frac{\omega(\Phi)}{\Phi^2}(\partial_{\mu} \Phi \partial_{\nu} \Phi
- \frac{1}{2}g_{\mu \nu}(\partial_{\alpha}\Phi)^2) \nonumber \\
&& + \frac{1}{\Phi} [\nabla_{\mu} \nabla_{\nu} -g_{\mu \nu}\Box]\Phi ,\label {eq:motiong}
\end{eqnarray}
and
\begin{equation}\label{eq:motionPhi}
\frac{2\omega(\Phi)+3}{\Phi}\Box \Phi= \frac{f(\Phi)}{\Phi} T - 2 f_{,\Phi}(\Phi) \mathcal{L}_m - \frac{\omega_{,\Phi}(\Phi)}{\Phi} (\partial_\sigma \Phi)^2 .
\end{equation}

\section{The $\gamma$ parameter and the 1PN/RM development}
\label{sec:magnitude}

In this section, we are interested in showing that the parameter $\gamma$ can take different values than usually expected. Therefore we develop the equations at the 1PN/RM order only \footnote{PN/RM stands for Post Newtonian/Relativistic Motion. It means that the development of the Post-Newtonian metric is developed to the order that has to be taken into account when dealing with test particles with relativistic velocities only. On the contrary, PN/SM stands for Post-Newtonian/Slow Motion. It means that the development of the Post-Newtonian metric is developed to the order that has to be taken into account when dealing with test particles with non-relativistic velocities only (for more details, see 2.1 in \cite{moiCQG11}).}. Let us write the perturbations of the fields  as follow:
\begin{eqnarray}
&&\Phi=\Phi_0+c^{-2} \varphi \label{eq:dev_phi}\\ 
&&g_{\mu \nu}=\eta_{\mu \nu}+ c^{-2} h_{\mu \nu}+O(c^{-3})\label{eq:dev_g},
\end{eqnarray}
where $\eta_{\mu \nu}$ is the metric of Minkowki and $\Phi_0$ is constant background field \footnote{For basic principles about post-Newtonian developments, see for instance \cite{damourBOOK87,DSX-I,Kopeikin_Vlasov_2004} and references therein. In \ref{sec:devSF}, we recall the reason why the perturbation of the scalar field can be developed using the same small parameter as with the metric.}.  Now, if one assumes the conservation of the matter fluid current ($\nabla_\sigma(\rho U^\sigma)=0$, where $c^2 \rho$ is the \textit{rest mass energy density} and $U^\alpha$ the four-velocity of the fluid), one has $\mathcal{L}_m=-\epsilon$, where $\epsilon$ is the \textit{total energy density} \cite{fockBOOK64,brownCQG93,bertolamiPRD08,harkoPRD10,moi_hPRD12}. Therefore, at the first order in the post-Newtonian development, one has $\mathcal{L}_m=-c^2 \rho + O(c^0)=T+O(c^0)$. Hence, equations (\ref{eq:motiong}) and (\ref{eq:motionPhi}) can be re-written at the first perturbative order as follows: 
\begin{eqnarray}
&&R^{\mu \nu}=\frac{f(\Phi_0)}{\Phi_0} \left(T^{\mu \nu}-\frac{1}{2} g^{\mu \nu} T \right) + \frac{1}{\Phi_0} \left(\partial^\mu \partial^\nu + \frac{1}{2} g^{\mu \nu} \triangle \right) \Phi \nonumber \\&&~~~~~~~~~~~+O(c^{-3}),\label{eq:Rmunu1}\\
&&\frac{2\omega_0 +3}{\Phi_0} \Box \Phi = \left(1+\Upsilon \right)\frac{f(\Phi_0)}{\Phi_0} T+O(c^{-3}), \label{eq:SF1PN/RM}
\end{eqnarray}
where $\omega_0 \equiv \omega(\Phi_0)$, and
\begin{equation}
\Upsilon \equiv  - 2 ~\Phi_0~ \frac{\partial \ln{f(\Phi)}}{\partial \Phi} |_{\Phi_0}.
\end{equation}
Defining 
\begin{eqnarray}
\sigma &\equiv & T^{00}/c^2+O(c^{-2}),\\
 G_{eff} &\equiv & \left(1 + \frac{1+\Upsilon}{2 \omega_0 +3} \right) \frac{c^4}{8 \pi} \frac{f(\Phi_0)}{\Phi_0},\\
 \gamma &\equiv & \frac{2 \omega_0 +2 - \Upsilon}{2 \omega_0 + 4 + \Upsilon},\label{eq:gammadef}
 \end{eqnarray}
the previous equations can be re-written as follows:
 \begin{eqnarray}
 &&R^{00}=c^{-2} \left\{ 4 \pi G_{eff} \sigma \right\}+O(c^{-3}),\\
 &&R^{ij}=c^{-2} \left\{-\delta_{ij} \gamma 4 \pi G_{eff} \sigma + \frac{1}{\Phi_0} \partial_i \partial_j \varphi \right\}+O(c^{-3})\\
 &&\frac{1}{\Phi_0} \triangle \varphi = - \frac{2+2\Upsilon}{2 \omega_0 + 4 + \Upsilon}  4 \pi G_{eff} \sigma+O(c^{-1}). \label{eq:phi2omeg}
 \end{eqnarray} 
 It is then straightforward to show that the metric solution can be put under the following standard PN/RM form 
 \begin{eqnarray}
 &&g_{00}=-1+c^{-2}\frac{2 w}{c^2}+O(c^{-3}),\label{eq:main_resulti}\\
 &&g_{0i}=O(c^{-3}),\\
 &&g_{ij}=\delta_{ij}\left(1+c^{-2}\frac{2 \gamma w}{c^2} \right)+O(c^{-3}),
 \end{eqnarray}
 where $\gamma$ is indeed a constant given by (\ref{eq:gammadef}), and where $w$ satisfies the equation of Newton at the first perturbative order :
 \begin{equation}
 \triangle w = -4 \pi G_{eff} \sigma +O(c^{-1}).\label{eq:main_resulte}
 \end{equation}
 The important fact to notice is that, depending on the value of $\Upsilon$ (and thus depending on the coupling function), $1-\gamma$ could be either positive, null \footnote{The author has recently became aware of the paper of Moffat and Toth \cite{moffatIJMPD12} in which they explored such a possibility in order to argue the possible solar system viability of Modified Gravity Theory (MOG) \cite{moffatJCAP06,moffatCQG09}.} or negative; while STT without coupling predict a positive value for finite value of $\omega_0$. 

This result is generalized to the most general parametrization in section \ref{sec:genSF} and is re-derived using the Einstein representation in \ref{sec:EF}.

\subsection{The conservation equation}

Because of the scalar/matter coupling, there is an energy transfer between the scalar-field and the material fields such that the conservation equation writes:
\begin{equation}
\nabla_\sigma \left[ f(\Phi) T^{\mu \sigma} \right]= \mathcal{L}_m f_{,\phi}(\Phi) \partial^\mu \Phi. \label{eq:conserv}
\end{equation}
According to equation (\ref{eq:conserv}), a point particle in a gravitational field only, is not following an inertial movement anymore because it does not follow a space-time geodesic in general (although photons do follow geodesics in this class of theories (see \ref{sec:geodesic}.). Also, one can verify that (\ref{eq:conserv}) is in accordance with \cite{harkoPRD13} in the appropriate limit.

However, let us notice that \cite{DamPolyGRG94,DamPolyNPB94} propose a decoupling mechanism such that the scalar/matter coupling could be driven toward a quasi-null coupling during the evolution of the Universe that would satisfy solar-system constraints on gravitation -- \cite{damourPRD10} even give quantitative predictions for possible violations of the equivalence principle. And it has to be noticed that such an action satisfies the condition for the dynamical decoupling studied in \cite{DamPolyGRG94,DamPolyNPB94} as long as $f(\Phi_{min})$ is a local minimum of $f(\Phi)$ when $\Phi_{min}$ is a local minimum of $\Phi$.

Also, let us notice that if $\Upsilon=-1$, then the scalar field decouples to the gravitating sources at the first order (see equation (\ref{eq:SF1PN/RM}) or later, equation (\ref{eq:SFERresc})) and therefore the deviation from the geodesic  is sent to the next order in the post-Newtonian development ($\Upsilon=-1 \Rightarrow \nabla_\sigma T^{\mu \sigma}  \sim 0$, since the perturbation of the scalar-field $\varphi$ is null at the first order).

\subsection{Remark regarding theories with extra dimensions}
\label{sec:xtradim}

In most cases, compactified extra dimensions imply $\omega_0(=\omega)$ to be of the order of $-1$ \cite{fujiiBOOKst,Sstring88}. Hence, in order to pass solar system tests that give $\omega_0 \gtrsim 10^4$ for non-massive scalar-fields \cite{PeriPRD10}, extra-dimensional theories have to imply a sufficient mass for the effective scalar field. Indeed, a very massive scalar-field has a frozen spatial dynamics on solar system scales, such that every value of $\omega_0$ satisfy solar system observations \cite{PeriPRD10} ($|1-\gamma| \lesssim 10^{-4}$ \cite{bertottiNATURE03}). However, in \cite{sensei07} is described a special case allowing to get rid of this condition, since a particular way of compactifying extra-dimensions leads to arbitrary $\omega(\Phi)$. 

As we can see, the universal scalar/matter coupling may allow to get rid of the massive scalar-field condition as well. Indeed, as long as $\Upsilon=-1+\mu$, with $|\mu|$ small enough, the theory could pass the solar system tests on $\gamma$, regardless the value of $\omega_0$. 

Therefore, there may exist a subclass of theories with compactified dimensions for which a decoupling of the scalar-field in the post-Newtonian regime naturally occurs. Hence, such theories naturally satisfy the strongest solar system's constraints; without requiring a large mass for the scalar-field (see section \ref{sec:magnitude2} for more details).

\subsection{The string dilaton case}
\label{sec:string}

 Let us remind that the action (\ref{eq:action}) is a generalization of the low-energy action predicted by string theories at tree-level (see equation (1) in \cite{DamPolyGRG94}); and a special case of the assumed action after full string loop expansion (see the second action in \cite{DamPolyNPB94}).  

As one can see, \cite{DamPolyGRG94,DamPolyNPB94} do not predict that the post-Newtonian constant $\gamma$ could be more or equal than one; while the main result of the present paper is to show that the scalar/matter coupling implies that $\gamma$ could be exactly equal to one, or be either less or more than one -- depending on the coupling function $f(\Phi)$. However, \cite{DamPolyGRG94,DamPolyNPB94} based their results on a non-interactive point-particle formalism; while the present paper is based on a perfect fluid formalism. Therefore, at a first glance, it seems that the two formalisms lead to different results. 

However, it turns out that \cite{DamPolyGRG94,DamPolyNPB94} have a mistake in the definition of a coupling parameter. This mistake leads to the apparent difference of results between the two formalisms.

\cite{DamPolyGRG94,DamPolyNPB94} work in the Einstein representation (also known as the Einstein \textit{frame}) such that, with the notations of the present paper, their action writes:
\be
S_{DamPoly}= \int d^4x \sqrt{-\ut{g}} \left(\frac{1}{4q} \tilde{R} - \frac{1}{2q} (\nabla \ul{\varphi})^2 \right) - \sum_\textrm{particles} \int \tilde{m}(\ul{\varphi})c d\tilde{s},
\ee
where $q$ is a coupling constant, $\ut{m}$ is the mass of particles in the Einstein representation and $\ut{g}_{\alpha \beta}$ is the metric in the Einstein representation -- related to the original representation by the conformal scalar $B_g$ through $\ut{g}_{\alpha \beta}=C B_g ~g_{\alpha \beta}$, where $C$ is some numerical constant. The resulting equations of motion write:
\bea
\ut{R}_{\mu \nu}= 2 \partial_\mu \ul{\varphi} \partial_\nu \ul{\varphi}+2q\left(\ut{T}_{\mu \nu}- \frac{1}{2} g_{\mu \nu}\ut{T} \right),\label{eq:Reinstein}\\
\Box \ut{\varphi}= -q  \alpha \ut{T}, \label{eq:faussescalar}
\eea
when $\alpha$ is defined as $\alpha=\partial \ln \ut{m} / \partial \ul{\varphi}$ and where we considered only one gravitational source (one particle)  in order to simplify the notations \footnote{Indeed, the paper deals with notations introduced in more than two papers and might become unnecessary difficult to follow without this simplification -- that does not change the discussion otherwise.}. Then, \cite{DamPolyGRG94,DamPolyNPB94} use an equation given in \cite{damourCQG92} that gives the parameter $\gamma$ as a function of the coupling parameter $\alpha$. The equation reads
\be
\gamma-1=- \frac{2 \alpha^2}{1+\alpha^2}|_{_0}, \label{eq:gammagiven}
\ee
such that $\gamma<1$ for finite real value of $\alpha |_{\Phi_0}$. The important point to notice is that in (2.7d) in \cite{damourCQG92}, $\alpha$ is defined as $\alpha=\partial \ln A / \partial \ul{\varphi}$ -- where $A$ is the square-root of the conformal factor given by $\ut{g}_{\alpha \beta}=A^{-2}(\ul{\varphi}) g_{\alpha \beta}$. Thus, identifying the definitions used in \cite{damourCQG92} and in \cite{DamPolyNPB94}, one has $B_g = A^{-2}$ -- and identifying with the notations of the current paper, one has $\Phi=B_g=A^{-2}$. On the other hand in \cite{DamPolyGRG94,DamPolyNPB94}, $\alpha$ is defined as $\alpha=\partial \ln \ut{m} / \partial \ul{\varphi}$, where $\ut{m}$ is the mass of the particle in the Einstein representation. In usual Brans-Dicke-like theories (ie. when $f(\Phi)$ is a constant) $\partial \ln A / \partial \ul{\varphi} = \partial \ln \ut{m} / \partial \ul{\varphi}$ since in that case one simply has $\ut{m} = A~ m$, where $m$ is the constant mass of the particle in the Jordan representation. However, the equality does not hold in the general case when $f(\Phi)$ is not a constant, and one has  $\partial \ln A / \partial \ul{\varphi} \neq \partial \ln \ut{m} / \partial \ul{\varphi}$ in general. Indeed, in general one has:
\bea
\frac{\partial \ln \ut{m}(A(\varphi),f(\varphi))}{\partial \varphi}&=&\frac{\partial \ln \ut{m}(A(\varphi),f(\varphi))}{\partial A} \frac{\partial A}{\partial \varphi}+\frac{\partial \ln \ut{m}(A(\varphi),f(\varphi))}{\partial f} \frac{\partial f}{\partial \varphi}. \label{eq:boom}
\eea
And because $f(\varphi)$ is in general independent to $A(\varphi)$, the last terms in (\ref{eq:boom}) shows that $\partial \ln A / \partial \varphi \neq \partial \ln \ut{m} / \partial \varphi$ in general. Now in particular, let us notice that \cite{DamPolyNPB94} assume that
\be
\ut{m}= \mu B_g^{-1/2} e^{-8 \pi^2 \nu B_g} \Lambda, \label{eq:assumpt}
\ee
where $\mu$ and $\nu$ are pure number of the order of unity and $\Lambda$ is the string cut-off mass scale. Since one has $B_g = A^{-2}$, one has $\partial \ln A / \partial \ul{\varphi} \neq \partial \ln \ut{m} / \partial \ul{\varphi}$. Therefore using equation (\ref{eq:gammagiven}) is not appropriate in the context considered by \cite{DamPolyGRG94,DamPolyNPB94,damourPRD10} -- even if the assumption (\ref{eq:assumpt}) was correct. 

Now, as in \ref{sec:RSSF}, let us define $\alpha_0=\partial \ln A / \partial \ul{\varphi} |_{\Phi_0}$, and the coupling strength $\alpha_2$ by:
\be
\Box \ul{\varphi}= -q  \alpha_2 \ut{T}. \label{eq:phiA2}
\ee
According to the previous discussion $\alpha_0 \neq \alpha_2$ in general. The conformal transformation of the Einstein metric to the metric in the original frame involves the transformation $g_{\alpha \beta}=A^2~\tilde{g}_{\alpha \beta}=[A_0^2+ 2 c^{-2} ~A_0 (\partial A/\partial \ul{\varphi})_0 ~\delta \ul{\varphi}+O(c^{-4})]~\tilde{g}_{\alpha \beta}$, where $\ul{\varphi}=\ul{\varphi}_0+c^{-2} \delta \ul{\varphi}$. Let us consider $A_0=1$ -- that simply means that one keeps the same metric's units in the two representations at the present epoch and does not restrict the generality -- one has $A_0(\partial A/\partial \ul{\varphi})_0 = \alpha_0$ and therefore $g_{\alpha \beta}= [1+ 2 c^{-2}~ \alpha_0 ~\delta \ul{\varphi}+O(c^{-4})]~\tilde{g}_{\alpha \beta}$.  Now, from equation (\ref{eq:Reinstein}), one deduces:
\be
c^{-2}\triangle \ut{w}= - q \ut{T}^{00} + O(c^{-4}),
\ee
where $\tilde{w}$ is the scalar potential of the Einstein metric. Therefore, from (\ref{eq:phiA2}), one deduces:
\be
\delta \ul{\varphi}= -  \alpha_2 \ut{w}+O(c^{-2}).
\ee
Hence one gets:
\be
g_{\alpha \beta}= \left[1-2c^{-2}~ \alpha_0 \alpha_2~ \tilde{w}+O(c^{-4})\right] ~\tilde{g}_{\alpha \beta}.
\ee
Developing the Einstein metric (that is such that it satisfies the so-called Strong Spatial Isotropic Condition (SSIC) -- ie. $\tilde{g}_{ij} \tilde{g}_{00} = - \delta_{ij} + O(c^{-4})$ \footnote{It has to be noticed that from (\ref{eq:Reinstein}) and (\ref{eq:faussescalar}), one gets $\ut{R}_{ij} - 1/2 \ut{g}_{ij} \ut{R}=O(c^{-4})$. Therefore, one can algebraically deduce that the Einstein metric satisfies the SSIC. For the derivation of this algebraic result, see \cite{DSX-I}.}), one gets the following equation for $\gamma$: 
\be
\gamma-1= - \frac{2 \alpha_0 \alpha_2}{1+ \alpha_0 \alpha_2}.
\ee
Hence, remembering that from solar system constraints one has $|\alpha_0| \sim |\alpha_2| \ll 1$, $\gamma-1$ can be positive if $\textrm{sign}(\alpha_0)=-\textrm{sign}(\alpha_2)$. Now, as demonstrated in \ref{sec:RSSF}, $\alpha_2=(1+\Upsilon) \alpha_0$ and therefore the equation for $\gamma$ results to:
\be
\gamma-1= - \frac{2 (1+\Upsilon) \alpha_0^2}{1+ (1+\Upsilon)\alpha_0^2},
\ee
which corresponds to the result given by equation (\ref{eq:gammadef}) (because $\alpha_0^{-2}= 2 \omega_0+3$). In particular, one recovers the fact that $\gamma>1$ for $\Upsilon<-1$. Moreover, let us notice that from (\ref{eq:genparang}) and (\ref{eq:genparanu}), one deduces that the special case considered in \cite{DamPolyGRG94} (that is the low energy action of string theories at tree-level) leads to $\gamma>1$. 

\subsection{Remark on current constraints coming form propagation of light observations}

As shown in \ref{sec:geodesic}, the geometric optic limit of the modified Maxwell equations leads to the usual geodesic equation for the propagation of light. Therefore, since the geodesic equation is the same as in general relativity minimally coupled to electromagnetism and since the metric can be written in the standard post-Newtonian form, one can use the usual constraints on the parameter $\gamma$ -- which have been obtained while assuming that space-time was accurately described by the standard PN metric in addition to the assumption that light was following space-time geodesics. Hence one gets from (\ref{eq:gammadef}):
\begin{equation}
\left\vert \frac{1+\Upsilon}{2\omega_0+3}\right\vert \lesssim \frac{|1-\gamma_{obs}|}{2},
\end{equation}
where $\gamma_{obs}$ is the value given by current observational constraints on the PN parameter $\gamma$. Let us note that the particular case of $\Upsilon=-1$ is very interesting because it passes the solar-system tests on $\gamma$, regardless the value of $\omega_0$. Thus, a wider range of theories seems to be viable than one would naively assume by considering previous results on Brans-Dicke-like scalar-tensor theories.

 Therefore, the lower limit deduced from observations for $\omega_0$, that is given in the context of usual Brans-Dicke-like scalar-tensor theories ($\omega_0 \sim |1-\gamma|^{-1}$), is not valid anymore in the context considered in this paper.

\section{The decoupling scenario: $f(\Phi) \propto \sqrt{\Phi}$}
\label{sec:magnitude2}

Contrary to the previous section, in this section we shall develop the equations at the 1.5PN order. Indeed, the case  $f(\Phi) \propto \sqrt{\Phi}$ -- a special case leading to $\Upsilon=-1$ -- is interesting because it leads to $\gamma=1$ and $\beta=1$ regardless the value of $\omega_0$. However, one has to quantify how much the theory deviates from general relativity at the post-Newtonian level, and compare with current observations in order to possibly get some constraints on $\omega_0$. Also, let us stress out that $f(\Phi) \propto \sqrt{\Phi}$ also implies that the theory satisfies the condition for the dynamical decoupling studied in \cite{DamPolyGRG94,DamPolyNPB94} since $\sqrt{\Phi_{min}}$ is a local minimum of $\sqrt{\Phi}$.

When $f(\Phi) \propto \sqrt{\Phi}$, one has $2f_{,\Phi}(\Phi)=\frac{f(\Phi)}{\Phi}$ and equation (\ref{eq:motionPhi}) writes:
\begin{equation}\label{eq:motionPhideco}
\frac{2\omega(\Phi)+3}{\Phi}\Box \Phi= \frac{f(\Phi)}{\Phi} \left(T -  \mathcal{L}_m\right) - \frac{\omega_{,\Phi}(\Phi)}{\Phi} (\partial_\sigma \Phi)^2 .
\end{equation}
Since $ \mathcal{L}_m=T+O(c^{0})$, the scalar field decouples to the gravitational sources at the Newtonian level (ie. $(2\omega(\Phi)+3) \frac{\Box \Phi}{\Phi}= - \frac{\omega_{,\Phi}(\Phi)}{\Phi} (\partial_\sigma \Phi)^2+O(c^{-4})$). Therefore, we shall develop the scalar field as follows: 
\begin{equation}
\Phi=\Phi_0+c^{-4} \phi.
\end{equation} 
With the following definition for the stress-energy tensor that is in accordance with $\mathcal{L}_m=-\epsilon$ \cite{moi_hPRD12}:
\be
T^{\mu \nu}= \left(\epsilon+P \right) U^\mu U^\nu + P g^{\mu \nu},
\ee
where $U^\alpha= c^{-1} dx^\alpha/d\tau$, P is the pressure of the fluid and $\epsilon$ its total energy density; the trace of the stress-energy tensor is $T=-\epsilon+3P$ and the the scalar field equation writes:
\be
\frac{2\omega(\Phi)+3}{\Phi}\Box \Phi= \frac{f(\Phi)}{\Phi} ~3P - \frac{\omega_{,\Phi}(\Phi)}{\Phi} (\partial_\sigma \Phi)^2 , 
\ee
which reduces to:
\be
\frac{\triangle \phi}{\Phi_0} = \frac{c^4 f(\Phi_0)}{\Phi_0} \frac{3 P}{2 \omega_0+3} + O(c^{-2}). \label{eq:phic-4}
\ee
Therefore, the time-time component of the Ricci tensor writes:
\be
R^{00}= \frac{1}{2} \frac{f(\Phi_0)}{\Phi_0} \left[\epsilon + \left(2 w + 2 v^2 \right) \frac{\epsilon}{c^2}+ \frac{2 \omega_0+2}{2 \omega_0+3}~ 3P \right]+ O(c^{-6}), \label{eq:r00dev}
\ee
where $v^2$ is the modulus squared of the coordinate velocity of the fluid. One can check that a PN metric with $\beta=1$ is an admissible solution of equation (\ref{eq:r00dev}). Hence, one can put the metric in a system of coordinate that would satisfy the Strong Isotropy Condition (ie. $\gamma=1$ and $\beta=1$). Then, one can algebraically develop the time-time component of the Ricci tensor in terms of the metric components as follows:
\bea
R^{00} &=& c^{-2} \left\{- \Box w \right\} + c^{-4} \left\{-4~ \partial_t H \right\}+O(c^{-6}),\\
&&\textrm{or} \nonumber \\
R^{00} &=& c^{-2} \left\{- \triangle w \right\} + c^{-4} \left\{-4~ \partial_t W \right\}+O(c^{-6}),
\eea
where $H \equiv \partial_t w + \partial_k w_k$ or $W \equiv 3/4~ \partial_t w + \partial_k w_k$ \cite{moiPRD09}\footnote{One has defined $g_{0i}=-4 c^{-3} w_i$ \cite{moiPRD09}.}. The harmonic gauge corresponds to $H=0$ and the standard post-Newtonian gauge corresponds to $W=0$. As an example, for W=0 one gets:
\be
\triangle w = -\frac{c^4}{2} \frac{f(\Phi_0)}{\Phi_0} \left[ \rho + c^{-2} \left\{\rho ~\Pi + \left(2 w + 2v^2 \right) \epsilon + \frac{2 \omega_0 +2}{2 \omega_0 +3} ~3P \right\} \right]+O(c^{-4}),
\ee
where $\Pi$ is the elastic compression potential energy per unit mass of the fluid \cite{fockBOOK64}. However, for both $W=0$ and $H=0$, one has:
\bea
w=  w_{GR} + c^{-2} \delta w,
\eea
where 
\be
\delta w =- \frac{3G_{eff}}{2\omega_0+3} \int  \frac{P(\bold{x}')d^3 x'}{|\bold{x}-\bold{x}'|} +O(c^{-4}).
\ee
Moreover, from (\ref{eq:phic-4}), one has:
\be
\frac{\phi}{\Phi_0}= 2 \delta w + O(c^{-2}).
\ee
Finally, let us stress out that the \textit{frame dragging potential} is unchanged at the 1.5PN level compared to general relativity:
\bea
w^i=  w^i_{GR} +O(c^{-2}).
\eea

\subsection{Massive test particles}
\label{sec:magnitude2a} 

Massive point-particles don't follow geodesics since the conservation equation writes:
\be
\nabla_\sigma  T^{\mu \sigma} = \frac{1}{2}\left( \mathcal{L}_m g^{\mu \sigma} - T^{\mu \sigma} \right) \frac{\partial_\sigma \Phi}{\Phi}.
\ee
Indeed, let us consider a flow of non-interactive massive point particles such that $T^{\alpha \beta}=c^2 \rho~ U^\alpha U^\beta$, where $U^\alpha = c^{-1} \d x^\alpha/ \d\tau$. Using the conservation of the matter fluid current $\nabla_\sigma (\rho U^\sigma)=0$ -- that is only valid for $\mathcal{L}_m=-\epsilon$ \cite{moi_hPRD12} --, one gets:
\be
U^\sigma \nabla_\sigma U^\mu = - \frac{1}{2} \left(g^{\mu \sigma}+U^\mu U^\sigma \right) \frac{\partial_\sigma \Phi}{\Phi}.
\ee
It means that massive point particles are not inertial in this class of theories. The correction to their trajectories thus comes from both the modification of the metric and the non-inertial acceleration $\vec{a}_{NI}$ ($a^i_{NI}\equiv - \frac{1}{2} \frac{\partial_i \Phi}{\Phi}$). Taking into account both of the correction, and using (\ref{eq:phic-4}), the final 1.5PN/SM equation in the time-coordinate parametrization reads:
\bea
\frac{\d^2 x^i}{c^2 \d t}+ \left(\digamma^i_{\alpha \beta}-\digamma^0_{\alpha \beta} \frac{\d x^i}{c \d t} \right) \frac{\d x^\alpha}{c \d t} \frac{\d x^\beta}{c \d t}&=&c^{-4} \left\{ \partial_i \delta w - \frac{1}{2} \frac{\partial_i \phi}{\Phi_0} \right\}+O(c^{-6})\nonumber \\&=&O(c^{-6}), \label{eq:trajMP}
\eea
where $\digamma^\gamma_{\alpha \beta}$ is the connection of general relativity.  Therefore, there is an exact cancellation between the non-inertial acceleration $\vec{a}_{NI}$ and the part of the inertial acceleration coming from the modification of the metric, and thus the trajectories of massive test particles are the same as in general relativity at the 1.5PN level -- even though they are not inertial anymore.

Therefore it seems that these theories cannot be differentiated from general relativity at the post-Newtonian level -- regardless the value of $\omega_0$. However, since the metric is modified from general relativity at the 1PN level through $\delta w$, the gravitational redshift is modified accordingly. But the relative deviation from general relativity is at best of the order of $\langle P/ (c^2 \rho) \rangle$ ($ \sim 10^{-6}$ for the Earth when assuming a mean pressure around 100 GPa). Future space experiments such as ACES \cite{cacciapuotiNPB07} and STE-QUEST \cite{cacciapuotiCOSPAR12} should be able to check the gravitational redshift at this accuracy.

\subsection{Photons}
\label{sec:magnitude2b} 

As seen in \ref{sec:geodesic}, photons still follow space-time geodesics. Therefore the 1.5 PN/RM equation for the trajectory of light is the same as in general relativity since the metric correction appears at the $c^{-4}$ level only -- and therefore does not contribute to the trajectory of light at the 1.5PN/RM level ($c^{-3}$ level). However, we expect that the deviation from general relativity will impact the photons' trajectories at the 2PN/RM level ($c^{-4}$ level). But current experiments are still far from being able to observe the 2PN/RM effects on the trajectory of light (see \cite{moiCQG11,dengPRD12} for instance).

Therefore, we conclude that the class of theories where $f(\Phi) \propto \sqrt{\Phi}$  is not constrained by current experiments in the solar system.

%
%

\section{Conclusion and final remarks}
\label{sec:concl}

In this paper we have shown that a universal scalar/matter coupling modifies the usual expression of the post-Newtonian parameter $\gamma$ in such a way that $1-\gamma$ could be either positive, null or negative for finite value of $\omega_0$; while it is usually thought to be positive only. In particular, we pointed out that previous studies considering similar couplings have missed that fact and we gave the reason for the apparent discrepancy. Moreover, it has to be stressed out that contrary to previous studies, the present paper did not use an assumption on the functional dependency of the particles' mass in the Einstein representation.

Also, we have focused our attention on a subclass of models that leads to a partial decoupling of the scalar-field to the gravitational sources at the post-Newtonian level; therefore leading to theories that are naturally closer to general relativity in the post-Newtonian regime than theories with general scalar/matter couplings. We showed that this class of models cannot be distinguish from general relativity for the trajectory of massive particles, as well as for experiment involving electromagnetic links such as the Cassini experiment \cite{bertottiNATURE03}. However, because the metric is slightly different from general relativity, we argued that near future gravitational redshift experiments such as ACES or STE/QUEST will be able to constrained the theory. 

Finally, let us note that one can expect the decoupling mechanism studied in section \ref{sec:magnitude2} to be greatly reduced in strong regimes where one may have $P \rightarrow c^2 \rho$. Therefore, strong regimes phenomena may give better constraints on the subclass of \textit{decoupled theories} than solar system experiments. However, this question have to be studied with cautious in order to be definitively answered.

\appendix 

\section{Development of the scalar field}
\label{sec:devSF}

The assumption that the scalar-field perturbation can be developed with the same small parameter as with the metric is justified by the field equations. Indeed, the sources of the two field equations are both proportional to the matter density: $\mathcal{L}_m \sim -c^2 \rho$ and $ \sim T \sim - T^{00} \sim -c^2 \rho$ -- the first equality is demonstrated in \cite{moi_hPRD12}; while the others come from the post-Newtonian assumptions as explained in \cite{DSX-I}. Therefore, unless $\Phi f_{,\Phi}/f$ is big, the relative perturbation of the scalar field is of the order of the relative perturbation of the metric. However, the PN parameter $\gamma$ is already measured to be very close to 1. Therefore, the relative perturbation of scalar field is necessarily much smaller than the relative perturbation of the metric and one does not have $\Phi f_{,\Phi}/f$  big in general -- at least in the solar system's neighborhood. 

Now since the order of magnitude of the relative perturbation of the scalar field is at best of the order of the relative perturbation of the metric, one can parametrize the development of the scalar field with the same parameter as with the metric. 

Let us stress that it is the usual procedure in post-Newtonian developments of alternative theories of gravitation (see, for instance, \cite{Will_book93,Kopeikin_Vlasov_2004,moiCQG12,dengPRD12}).

\section{The geometric optic limit}
\label{sec:geodesic}

Let's consider the geometric optic limit of Electromagnetism in the theory studied here. The electromagnetic field amplitude considered being extremely weak -- a laser of a few Watts for instance -- the electromagnetic field won't affect the field equations previously considered (ie. photons are considered as test particles). However, the Maxwell equation in vacuum is modified by the scalar field in the following way:
\begin{equation}
\nabla_\sigma \left(f(\Phi) F^{\mu \sigma} \right)=0.
\end{equation}
Using the Lorenz Gauge ($\nabla_\sigma A^\sigma =0$), along with equation (\ref{eq:phi2omeg}), one puts this equation under the 1-PN/RM following form:
\begin{equation}
-\Box A^\mu + g^{\mu \epsilon} R_{\gamma \epsilon} A^\gamma + \varkappa \left(\nabla^\mu A^\sigma - \nabla^\sigma A^\nu \right) \partial_\sigma w=O(c^{-3}), \label{eq:maxmodLG}
\end{equation}
where
\begin{equation}
\varkappa \equiv c^{-2} \Phi_0 \frac{f_{,\Phi}}{f}|_{\Phi_0} \frac{2+2\Upsilon}{2 \omega+4+\Upsilon}.
\end{equation}
Following the analysis made in \cite{MTW}, we expand the 4-vector potential as follows:
\begin{equation}
A^\mu = \Re \left\{ \left(a^\mu + \epsilon b^\mu + O(\epsilon^2) \right) \exp^{i \theta / \epsilon} \right\},
\end{equation}
The two first leading orders of equation (\ref{eq:maxmodLG}) respectively give:
\begin{equation}
k_\sigma k^\sigma =O(c^{-3}), 
\end{equation}
where $k_\sigma \equiv \partial_\sigma \theta$, and
\begin{equation}
a^\mu \nabla_\sigma k^\sigma + 2 k^\sigma \nabla_\sigma a^\mu + \varkappa \left(k^\mu a^\sigma - k^\sigma a^\mu \right) \partial_\sigma w = O(c^{-3}).
\end{equation}
Remembering that the Lorenz Gauge condition gives $k_\sigma a^\sigma=0$ at the leading order, one gets:
\begin{equation}
k^\sigma \nabla_\sigma k^\mu = O(c^{-3}).
\end{equation}
This equation is the usual geodesic equation, showing that the presence of the scalar-field won't affect light ray trajectories at the geometric optic approximation. However, defining $a^\mu = a f^\mu$, the propagation equation for the scalar amplitude ($a$) as well as the propagation equation for the polarization vector ($f^\mu$) are modified:
\begin{eqnarray}
&&k^\sigma \nabla_\sigma a = - \frac{a}{2} \nabla_\sigma k^\sigma + \frac{\varkappa}{2} a k^\sigma \partial_\sigma w+O(c^{-3}), \\
&&k^\sigma \nabla_\sigma f^\mu = + \frac{\varkappa}{2} k^\mu f^\sigma \partial_\sigma w+O(c^{-3}).
\end{eqnarray}
From there follows that the conservation law of "photon number" is modified:
\begin{equation}
\nabla_\sigma (k^\sigma a^2)= -\varkappa a^2 k^\sigma \partial_\sigma w+O(c^{-3}).
\end{equation}
One notes that the last three equations give alternative ways to put constraints on those theories. Those ways should be investigate using the relevant literature.\\

Otherwise, one should notice that $\Upsilon=-1$ implies $\varkappa=0$ in addition to $\gamma=1$. Meaning that the photon number is conserved at the 1.5PN/RM level. However let us stress that, even in the case $\Upsilon=-1$, we expect a violation of the conservation at the 2PN/RM level.

\section{General parametrization of the scalar field}
\label{sec:genSF}

If, instead of action (\ref{eq:action}), one starts with the following general action:
\begin{eqnarray}
S=\int d^4x \sqrt{-g} \big\{&& F(\Phi)R-Z(\Phi) (\partial_\sigma \Phi)^2  + f(\Phi) \mathcal{L}_m (g_{\mu \nu}, \Psi) \big\},\label{eq:actiondila}
\end{eqnarray}
then the $\gamma$ parameter re-writes:
\begin{equation}
\gamma= {\frac{2 Z F+(2-\Upsilon)(F_{,\Phi})^2}{2 Z F+(4+\Upsilon)(F_{,\Phi})^2}} \big |_{\Phi_0}, \label{eq:genparang}
\end{equation}
with 
\begin{equation}
\Upsilon \equiv -2 {\frac{F}{F_{,\Phi}}} \big |_{\Phi_0} {\frac{f_{,\Phi}}{f}} \big |_{\Phi_0}.\label{eq:genparanu}
\end{equation}
The conservation equation still writes:
\begin{equation}
\nabla_\sigma \left[ f(\Phi) T^{\mu \sigma} \right]= \mathcal{L}_m f_{,\phi}(\Phi) \partial^\mu \Phi.
\end{equation}
Again, let us remark that $f \propto \sqrt{F}$ leads to $\gamma=1$ (as well as the conservation of the photon number at the 1.5PN/RM level ($\varkappa=0$ in \ref{sec:geodesic})) -- regardless the value of the kinetic function $Z$.

\section{Using the Einstein representation}
\label{sec:EF}

The results presented in this paper do not depend on the representation used to do the calculations. However, it is always interesting to re-derive the calculations in the Einstein representation in order to check the results obtained while using the original representation only. The action writes in the original and Einstein representation respectively as follows:
\begin{eqnarray}
S&=&\int d^4x\sqrt{-g} \left( \Phi R -
\frac{\omega(\Phi)}{\Phi} g^{\alpha \beta} \partial_\alpha \Phi \partial_\beta \Phi \right) + S_m,\\
&=& \int d^4x\sqrt{-\tilde{g}} \left(  \tilde{R} -
\left(\omega(\Phi(\varphi))+\frac{3}{2} \right) \tilde{g}^{\alpha \beta} \partial_\alpha \varphi \partial_\beta \varphi \right) + S_m, \nonumber \\ \label{eq:actionER}
\end{eqnarray}
where $g^{\alpha \beta} \equiv \Phi \tilde{g}^{\alpha \beta}$, $\sqrt{-g}= \Phi^{-2} \sqrt{-\tilde{g}} $ and $\varphi \equiv \ln \Phi$. By definition, the material part of the action ($S_m$) writes:
\begin{eqnarray}
S_m&=&\int d^4x\sqrt{-g}~ 2 f(\Phi) \mathcal{L}_m(g_{\mu \nu}, \Psi),\label{eq:actionMJ}\\
&=& \int d^4x\sqrt{-\tilde{g}} ~2 f(\Phi(\varphi)) \tilde{\mathcal{L}}_m(\ut{g}_{\nu \nu}, \Phi, \Psi).
\end{eqnarray}
Therefore, by definition, one has $\tilde{\mathcal{L}}_m=\Phi^{-2}\mathcal{L}_m$. 

The equations of motion given by the action in the Einstein representation are easily derived from (\ref{eq:actionER}). However, it is not trivial to figure out what is the source $\sigma$ of the scalar-field $\varphi$ in the Einstein representation, where $\sigma= (-\tilde{g})^{-1/2} \delta S_m/\delta {\varphi}$. For instance, \cite{DamPolyGRG94,DamPolyNPB94} use an assumption on the functional dependency of the Einstein mass $\ut{m}$  (\ref{eq:assumpt}); instead of deriving the dependency from the action in the original representation. In the following, we expound the derivation of $\sigma$. 

The variation of equation (\ref{eq:actionMJ}) for relevant fields leads to:
\begin{equation}
\delta S_m=\int d^4x\sqrt{-g} \left(-f(\Phi) T_{\alpha \beta} ~\delta g^{\alpha \beta}+2 f_{,\Phi}(\Phi) \mathcal{L}_m ~\delta \Phi \right).\label{eq:varSm}
\end{equation}
Now, since one has $g^{\alpha \beta} \equiv \Phi \tilde{g}^{\alpha \beta}$, the variation of the \textit{physical} metric gives
\begin{equation}
\delta g^{\alpha \beta}= \tilde{g}^{\alpha \beta} ~\delta \Phi + \Phi ~\delta \tilde{g}^{\alpha \beta}.
\end{equation}
Therefore, equation (\ref{eq:varSm}) writes:
\begin{eqnarray}
\delta S_m=\int d^4x\sqrt{-g} \left(-\Phi f(\Phi) T_{\alpha \beta} ~\delta \tilde{g}^{\alpha \beta} + \left[-f(\Phi) \tilde{g}^{\alpha \beta} T_{\alpha \beta}+ 2 f_{,\Phi}(\Phi) \mathcal{L}_m \right]~ \delta \Phi \right).\nonumber \\
\end{eqnarray}
Now, using $T_{\alpha \beta}= \Phi \tilde{T}_{\alpha \beta}$, $\tilde{T} \equiv \tilde{g}^{\alpha \beta} \tilde{T}_{\alpha \beta}$, $\delta \varphi = \delta \Phi / \Phi$ and $\Phi f_{,\Phi}(\Phi)=f_{,\varphi}(\Phi(\varphi))$, one gets:
\begin{eqnarray}
\delta S_m=&&\int d^4x\sqrt{-\tilde{g}} \big(- f(\Phi(\varphi)) \tilde{T}_{\alpha \beta}~ \delta \tilde{g}^{\alpha \beta} \label{eq:deltaSM} \\
&&- \left[1- 2 \frac{f_{,\varphi}(\Phi(\varphi))}{f(\Phi(\varphi))} \frac{\tilde{\mathcal{L}}_m}{\tilde{T}} \right]f(\Phi(\varphi)) ~\tilde{T}  ~\delta \varphi \big).\nonumber
\end{eqnarray}
The second part of the right hand side of (\ref{eq:deltaSM}) gives the sought-after $\sigma$. 

Now, since $\tilde{\mathcal{L}}_m=\Phi^{-2} \mathcal{L}_m$ and $\tilde{T}= \Phi^{-2} T$, $\tilde{\mathcal{L}}_m / \tilde{T}$ reduces to $\mathcal{L}_m / T=1+O(c^{-2})$ \cite{moi_hPRD12}. 
%

\subsection{Rescaling of the scalar-field, and correction of Damour and Polyakov's equation for $\gamma$}
\label{sec:RSSF}

While one can work with the action (\ref{eq:actionER}) in the Einstein representation, the scalar-field is often rescaled such that the action writes:
\begin{eqnarray}
S= \int d^4x\sqrt{-\tilde{g}} \left(  \tilde{R} - \tilde{g}^{\alpha \beta} \partial_\alpha \ul{\varphi} \partial_\beta \ul{\varphi} \right) + S_m, \label{eq:SEwRS}
\end{eqnarray}
where $\d \ul{\varphi}=\pm \sqrt{\omega + 3/2} ~\d \varphi$. In what follows we consider the \textit{re-scaled} action (\ref{eq:SEwRS}) only in order to compare our result with previous studies. Choosing the re-scaling $\d \ul{\varphi}= \sqrt{\omega + 3/2} ~\d \varphi$, one can re-write (\ref{eq:deltaSM}) as follows:
\begin{eqnarray}
\delta S_m=\int d^4x\sqrt{-\tilde{g}} \big(&&- f(\Phi(\ul{\varphi})) \tilde{T}_{\alpha \beta}~ \delta \tilde{g}^{\alpha \beta} \label{eq:deltaSM2} \\
&&- \left[1- 2 \frac{f_{,\varphi}(\Phi(\varphi))}{f(\Phi(\varphi))} \frac{\tilde{\mathcal{L}}_m}{\tilde{T}} \right] \frac{f(\Phi(\ul{\varphi}))}{\sqrt{\omega(\Phi(\ul{\varphi}))+3/2}} ~\tilde{T}  ~\delta \ul{\varphi} \big).\nonumber
\end{eqnarray}
Note that the second part of the right hand side gives the source $\ul{\sigma}$ of the scalar-field $\ul{\varphi}$, with $\ul{\sigma}= (-\tilde{g})^{-1/2} \delta S_m/\delta \ul{\varphi}$. From (\ref{eq:SEwRS}) and (\ref{eq:deltaSM2}), one gets the following 1.5PN/RM equation for $\ul{\varphi}$:
\begin{equation}
\triangle \ul{\varphi} = - \alpha_0 f(\Phi_0) (1+\Upsilon) \tilde{T}+O(c^{-4}), 
\end{equation}
with $\alpha$ defined in \cite{damourCQG92} by $\alpha \equiv \partial \ln{A}/ \partial \ul{\varphi}$ with $g_{\alpha \beta}= A^2(\ul{\varphi}) \tilde{g}_{\alpha \beta}=\Phi^{-1}\tilde{g}_{\alpha \beta}$. In order to compare with \cite{DamPolyGRG94,DamPolyNPB94}, let us write:
\begin{equation}
\triangle \ul{\varphi} =  - \alpha_2 ~f(\Phi_0)\tilde{T}+O(c^{-4}) , \label{eq:SFERresc}
\end{equation}
with 
\begin{equation}
\alpha_2 \equiv \alpha_0~ (1+\Upsilon). \label{eq:SFERrescalpha}
\end{equation}
Therefore, as suggested in section \ref{sec:string}, the coupling strength $\alpha_2$ is in general different from $\alpha_0=\partial \ln A / \partial \ul{\varphi}|_{\Phi_0}$. On the other hand, from (\ref{eq:SEwRS}) and (\ref{eq:deltaSM2}), the Newtonian potential in the Einstein representation satisfies:
\begin{equation}
c^{-2}~\triangle \tilde{w} = -  f(\Phi_0) \tilde{T}^{00}+O(c^{-4}). \label{eq:SFER}
\end{equation}
Therefore, one has $g_{\alpha \beta}=A^2~\tilde{g}_{\alpha \beta}= (1-2c^{-2}~ \alpha_0 \alpha_2~ \tilde{w}+O(c^{-4}))~ \tilde{g}_{\alpha \beta}$  \footnote{see section \ref{sec:string}.}. Now, remembering that the Einstein metric satisfies the strong spatial isotropy condition ($\tilde{g}_{ij} \tilde{g}_{00} = - \delta_{ij} + O(c^{-4})$), one gets for the PN parameter $\gamma$:
\begin{equation}
\gamma = \frac{1- \alpha_0 \alpha_2}{1+ \alpha_0  \alpha_2} =\frac{1- \alpha_0^2 (1+\Upsilon)}{1+ \alpha_0^2 (1+\Upsilon)}, \label{eq:gammaER}
\end{equation}
or
\begin{equation}
\ul{\gamma} = -\frac{2 \alpha_0^2 (1+\Upsilon)}{1+ \alpha_0^2 (1+\Upsilon)},
\end{equation}
where $\ul{\gamma} \equiv \gamma-1$.  Therefore, it shows that the parameter $\Upsilon$ is missing in the formula for the $\gamma$ parameter given in \cite{DamPolyGRG94,DamPolyNPB94,damourPRD10}. Accordingly, their parameter can only be less than one (see (9) in \cite{DamPolyGRG94} for instance); while we have shown that, depending on the coupling function, it could actually be either positive, null or negative. The discrepancy comes from the wrong assumption in \cite{DamPolyGRG94,DamPolyNPB94} that $\partial \ln A / \partial \ul{\varphi} = \partial \ln \ut{m} / \partial \ul{\varphi}$. Also, in (\ref{eq:SFERrescalpha})  we show how the coupling strength $\alpha_2$ defined in (\ref{eq:phiA2}) relates to $\alpha_0=\partial \ln A / \partial \ul{\varphi}|_{\Phi_0}$ in general.

Now, remembering that $\alpha_0^{-2} = 2\omega_0+3$, one exactly gets (\ref{eq:gammadef}) from (\ref{eq:gammaER}).

\ack
This research was partly supported by an appointment to the NASA Postdoctoral Program at the Jet Propulsion Laboratory, California Institute of Technology, administered by Oak Ridge Associated Universities through a contract with NASA. \copyright 2012 California Institute of Technology. Government sponsorship acknowledged. \\
This research was partly done as an invited researcher of the Observatoire de la C\^ote d'Azur.\\
The author wants to thank Tiberiu Harko, Viktor Toth, John Moffat and Aurelien Hees for interesting discussions and comments. 
\\\\
\bibliographystyle{jphysicsB}


\end{document}